\documentclass[longauth]{aa}  
\usepackage{graphicx}
\usepackage{txfonts}
\newcommand{\teff}{$T_{\rm eff}$}
\newcommand{\ewli}{${\rm EW}({\rm Li})$}
\newcommand{\dewli}{$\delta {\rm EW}({\rm Li})$}
\newcommand{\msun}{$M_\odot$}

\def\kms{km~s$^{-1}$}

\def\logg{$\log g$}
\def\feh{[Fe/H]}

\def\vsini{$v\sin i$}
\def\vrad{${\rm V}_{rad}$}

\def\wli{${\rm EW}({\rm Li})$}
\def\ha10{${\rm H}\alpha\,10\%$}

\def\li{Li\,(6707.84\,\AA)}

\begin{document}

   \title{The \textsl{Gaia}-ESO Survey: A lithium-rotation connection
      at 5 Myr?}


   \author{J. Bouvier \inst{1,2}
          \and
          A.~C. Lanzafame \inst{3,4}
          \and
          L. Venuti \inst{1,2,5}
          \and
          A. Klutsch \inst{4}
          \and
          R. Jeffries \inst{6}
          \and
          A. Frasca \inst{4}
\and
	E. Moraux \inst{1,2}
           \and
          K. Biazzo \inst{4}
          \and
          S. Messina \inst{4}
          \and
          G. Micela \inst{5}
                   \and
          S. Randich \inst{7}
          \and
          J. Stauffer \inst{8}
          \and
          A.~M. Cody \inst{9}
          \and
          E. Flaccomio \inst{5}
          \and
          G. Gilmore \inst{10}
          \and
          A. Bayo \inst{11}
          \and
          T. Bensby\inst{12}
          \and 
          A. Bragaglia \inst{13}   
          \and
          G. Carraro \inst{14}  
          \and
          A. Casey \inst{10}
          \and
          M.~T. Costado \inst{15}
          \and
          F. Damiani \inst{5}
          \and   
          E. Delgado Mena \inst{16}
          \and
          P. Donati \inst{13}
          \and
          E. Franciosini \inst{7}
          \and 
          A. Hourihane \inst{10}
          \and
          S. Koposov \inst{10}
          \and
          C. Lardo \inst{17}
          \and
          J. Lewis \inst{10}
          \and 
          L. Magrini \inst{7}
          \and
          L. Monaco \inst{18}
          \and
          L. Morbidelli \inst{7}
          \and
          L. Prisinzano \inst{5}
          \and
          G. Sacco \inst{7}
          \and
          L. Sbordone \inst{19}
          \and
          S.~G. Sousa \inst{16}
          \and
          A. Vallenari \inst{20}
          \and
          C.~C. Worley \inst{10}
          \and
          S. Zaggia \inst{20}
          \and
          T. Zwitter \inst{21}
          }




   \institute{Univ. Grenoble Alpes, IPAG, F-38000 Grenoble, France\\
     \email{Jerome.Bouvier@obs.ujf-grenoble.fr}
     \and
      CNRS, IPAG, F-38000 Grenoble, France 
         \and
          Universit\`a di Catania, Dipartimento di Fisica e Astronomia, Sezione
         Astrofisica, Via S. Sofia 78, I-95123 Catania, Italy\\
         \email{Alessandro.Lanzafame@oact.inaf.it}    
         \and
          INAF - Osservatorio Astrofisico di Catania, Via S. Sofia 78, I-95123 Catania, Italy
         \and
          INAF - Osservatorio Astronomico di Palermo G.S. Vaiana, Piazza del Parlamento 1, 90134 Palermo, Italy
         \and
         Astrophysics Group, Keele University, Keele, Staffordshire, ST5 5BG, UK
	         \and
	       INAF - Osservatorio Astrofisico di Arcetri, Largo E. Fermi 5, 50125,
Florence, Italy
	      \and   
	       Spitzer Science Center, California Institute of Technology, Pasadena, CA 91125, USA
	      \and
	       NASA Ames Research Center, Kepler Science Office, Mountain View, CA 94035, USA
	      \and
	       Institute of Astronomy, University of Cambridge, Madingley Road, Cambridge CB3 0HA, UK
	      \and
	       Instituto de F\'{\i}sica y Astronom\'{\i}a, Universidad de Valpara\'{\i}so, Chile
	      \and
	       Lund Observatory, Dept of Astronomy and Theoretical Physics, Box 43, 22100 Lund, Sweden
	      \and
	       INAF - Osservatorio Astronomico si Bologna, via Ranzani 1, 40127 Bologna, Italy
	      \and
	       European Southern Observatory, Alonso de Cordova 3107, Santiago, Chile
	      \and
	       Instituto de Astrofisica de Andaluc\'{\i}a -- CSIC, Apdo. 3004, 18008 Granada, Spain
	      \and
	       Instituto de Astrof\'isica e Ci\^encias do Espa\c{c}o, Universidade do Porto, CAUP, Rua das Estrelas, 4150-762 Porto, Portugal
	      \and
	       Astrophysics Research Institute, Liverpool John Moores University, 146 Brownlow Hill, Liverpool L3 5RF, UK
	      \and
	       Universidad Andres Bello, Departamento de Ciencias F\'{\i}sicas, Facultad de Ciencias Exactas, Fernandez Concha 700, Las Condes, Santiago, Chile
	      \and
	       Millennium Institute of Astrophysics, Pontificia Universidad Cat\'olica de Chile, Vicu\~na Mackenna 4860, Macul, Santiago, Chile
	      \and
	       INAF - Osservatorio Astronomico di Padova, Vicolo Osservatorio 5, 35122 Padova, Italy
	      \and
	       University of Ljubljana, Faculty of Mathematics and Physics, Jadranska 19, 1000 Ljubljana, Slovenia 
             }

   \date{Received February 18, 2016 ; accepted April 25, 2016}

 
  \abstract
   {The evolution of lithium abundance in cool dwarfs provides a unique probe of non-standard processes in stellar evolution.}
   {We investigate here the lithium content of young low-mass stars in the 5~Myr-old star forming region NGC~2264 and its relationship with rotation. 
  }
   {We combine lithium equivalent width measurements (\ewli) from the \textsl{Gaia}-ESO Survey with the determination of rotational periods from the CSI~2264 survey. We consider only {\it bona fide} non accreting cluster members in order to minimize uncertainties on \ewli.  }
   {We report the existence of a relationship between lithium content and rotation in NGC~2264 at an age of 5~Myr. The Li-rotation connection is seen over a restricted temperature range (\teff=3800--4400~K) where fast rotators are Li-rich compared to slow ones. This correlation is similar to, albeit of lower amplitude than, the Li-rotation connection previously reported for K dwarfs in the 125 Myr-old Pleiades cluster. We investigate whether the non-standard pre-main sequence models developed so far to explain the Pleiades results, which are based on episodic accretion, pre-main sequence core-envelope decoupling, and/or radius inflation due to enhanced magnetic activity, can account for an early development of the Li-rotation connection. While radius inflation appears to be the most promising possibility, each of these models has issues. We therefore also discuss external causes that might operate during the first few Myr of pre-main sequence evolution, such as planet engulfment and/or steady disk accretion, as possible candidates for the common origin for Li-excess and fast rotation in young low-mass pre-main sequence stars.}
   {The emergence of a connection between lithium content and rotation rate at such an early age as 5~Myr suggests a complex link between accretion processes, early angular momentum evolution, and possibly planet formation, which likely impacts early stellar evolution and still is to be fully deciphered. }


   \maketitle
%

\section{Introduction}

Lithium is a sensitive probe of stellar evolution and most importantly of non-standard transport processes occurring in the stellar interiors. A fragile element, lithium is burnt at a temperature of 2.5~MK, which corresponds to the temperature at the base of the convective zone of a solar-mass star on the zero-age main sequence \citep[ZAMS,][]{Siess00}. Hence, lithium is slowly depleted and its surface abundance steadily decreases over time in solar-type and lower mass stars \citep[e.g.][]{Basri91, Martin94, Sestito05, Randich10, Jeffries14}. The rate of this secular evolution may be modified, either increased or reduced, by so-called non-standard transport processes, such as rotational mixing, internal magnetic fields, gravity waves, or tidal interactions, as well as by structural changes induced by, e.g., rotation, magnetic activity, metallicity, or accretion  \citep[e.g.][]{Pinsonneault90, Zahn92, Ventura98, Piau02, Talon05, Denissenkov10, Eggenberger12, Theado12}.  These additional processes may vary from star to star, depending on initial conditions and specific evolutionary paths, and thus have the potential to produce an enhanced lithium spread in otherwise similar stars at the same age \citep[e.g.][]{Pasquini08,	doNascimento10, Ramirez12}. 

In an attempt to relate lithium scatter to other stellar properties, \cite{Soderblom93} reported a clear connection between lithium abundance and rotation rate among the 125 Myr-old Pleiades K dwarfs. The finding that fast rotators were systematically more Li-rich than their slowly rotating counterparts was the first clear empirical evidence that rotation could indeed impact on the lithium abundance of solar-type stars as early as at the ZAMS stage. This result actually came as a surprise as rotational mixing from dynamical instabilities was thought to scale with surface rotation, which would predict fast rotators to be more lithium-depleted compared to slow ones \citep[e.g.][]{Pinsonneault89}, i.e., a trend opposite to what was observed in the Pleiades. Since then, various ideas have been proposed that could potentially account for the Pleiades Li-rotation connection, including the impact of initial disk lifetimes on rotational mixing \citep{Bouvier08, Eggenberger12}, and   structural changes linked to magnetic activity \citep{Ventura98, Somers14, Somers15} or  episodic accretion \citep{Baraffe10}.  Weaker evidence for a lithium-rotation connection has  also been reported for ages slightly younger than the Pleiades, e.g., for the 80 Myr-old Alpha Per cluster \citep{Randich98, Balachandran11} but it is unclear at what age it begins to develop.

Tracing the evolving pattern of lithium abundance and its spread as a function of age offers a unique opportunity to gain more insight into the fundamental processes occuring in stellar interiors. Our goal here is to investigate whether the link between lithium content and rotation may already be present at an age much earlier than the ZAMS \citep[cf.][]{Martin94}.  We thus focus on the low-mass stellar population of the young open cluster NGC~2264, at an age of 3--6~Myr \citep{Mayne08, Gillen14}. In order to build a large and homogeneous sample of low-mass pre-main sequence (PMS) stars in this cluster that allows us to investigate the lithium-rotation connection at very young ages, we combined two major surveys. Namely we take advantage of: i) precise lithium measurements \citep[cf.][]{Lanzafame15} obtained with VLT/FLAMES for the low-mass members of NGC~2264 in the framework of the \textsl{Gaia}-ESO Survey \citep[GES,][]{Gilmore12, Randich13}, and ii) the derivation of accurate rotational periods (Venuti et al., in prep.; see also \citet{Affer13}) from the CoRoT light curves of the CSI~2264 campaign \citep[Coordinated Synoptic Investigation of NGC~2264,][]{Cody14} that photometrically monitored the NGC~2264 star forming region over 40 days.  In Sect.~2, we describe the sample we built from the combination of these two surveys. In Sect.~3, we investigate the connection between lithium  and rotation in this sample, and report a statistically significant relationship between lithium equivalent width (\ewli) and rotational period over a limited \teff\ range. In Sect.~4, we review how the various scenarios developed to reproduce the Pleiades lithium scatter may account for its emergence on a much shorter timescale during the early PMS, and consider as well external factors such as disk accretion and/or planet engulfment as alternatives. The conclusions of this study are given in Sect.~5. 

\section{The sample}

The sample we use in this study was built by cross-correlating the
CSI~2264 database (Cody et al. 2014) with the 4th internal data release (iDR4) of the GES. From the former we extracted photometry,
fundamental parameters (mass, radius) and rotational periods, and
from the latter \ewli, radial and rotational
velocities, and effective temperatures. Specifically, we  
first selected all stars from CSI~2264 classified either as 
weak-line (WTTS) or as classical T Tauri stars (CTTS) in \cite{Venuti14} that also had a measured photometric (rotational)
period, most of which come from Venuti et al. (in prep.) and some
from the literature \citep{Lamm04, Affer13}. We then
cross-correlated this subsample with the GES database, which 
yielded 
295 objects with a \teff\ less than 6600~K, of which 217 are WTTS. Of the latter, 16 lacked an \ewli\ value in the GES iDR4, even though most exhibited a strong lithium absorption line in their spectrum. Appendix~A provides details on \ewli\ measurements.  As explained there, the lack of internal agreement between \ewli\ values reported for these stars, mostly due to difficulties in estimating the continuum level, prevented a recommended value to be defined. We note that 12 of these sources are M-type stars, and 9 are fast rotators, with \vsini\ larger than 80~\kms, which might account for the lack of consistent \ewli\ measurements. Only two of those stars lie in the 3800--4400~K \teff\ range that we discuss in more details in Section 3 below\footnote{These are Mon 869 (\teff=3931 K, \vsini=12.9~\kms, and Mon 1254 (\teff=4231 K, \vsini=117.5~\kms).}, and are therefore unlikely to affect our results. We are thus left with 201 WTTS with known stellar parameters, rotational periods, and \ewli, whose properties are listed
in Table 1{\footnote{Even though the GES iDR4 release provides Lithium {\it abundances} for this subsample, we prefer to use here \ewli\ because uncertainties on the stellar parameters, mostly on \teff, do not propagate on \ewli\ (except for line blends, cf. Appendix~A), while they do on abundances. Moreover, \ewli\  
is directly measured on the spectra while abundances are model-dependent.}.  While most of these stars have a strong membership
likelihood, based on photometry, accretion diagnotics and/or X-ray emission \citep[see][]{Venuti14}, a few were rejected as non-members upon further analysis
(see Table 1 and next section).


\section{Results}

     \begin{figure}
   \centering
    \includegraphics[width=9cm]{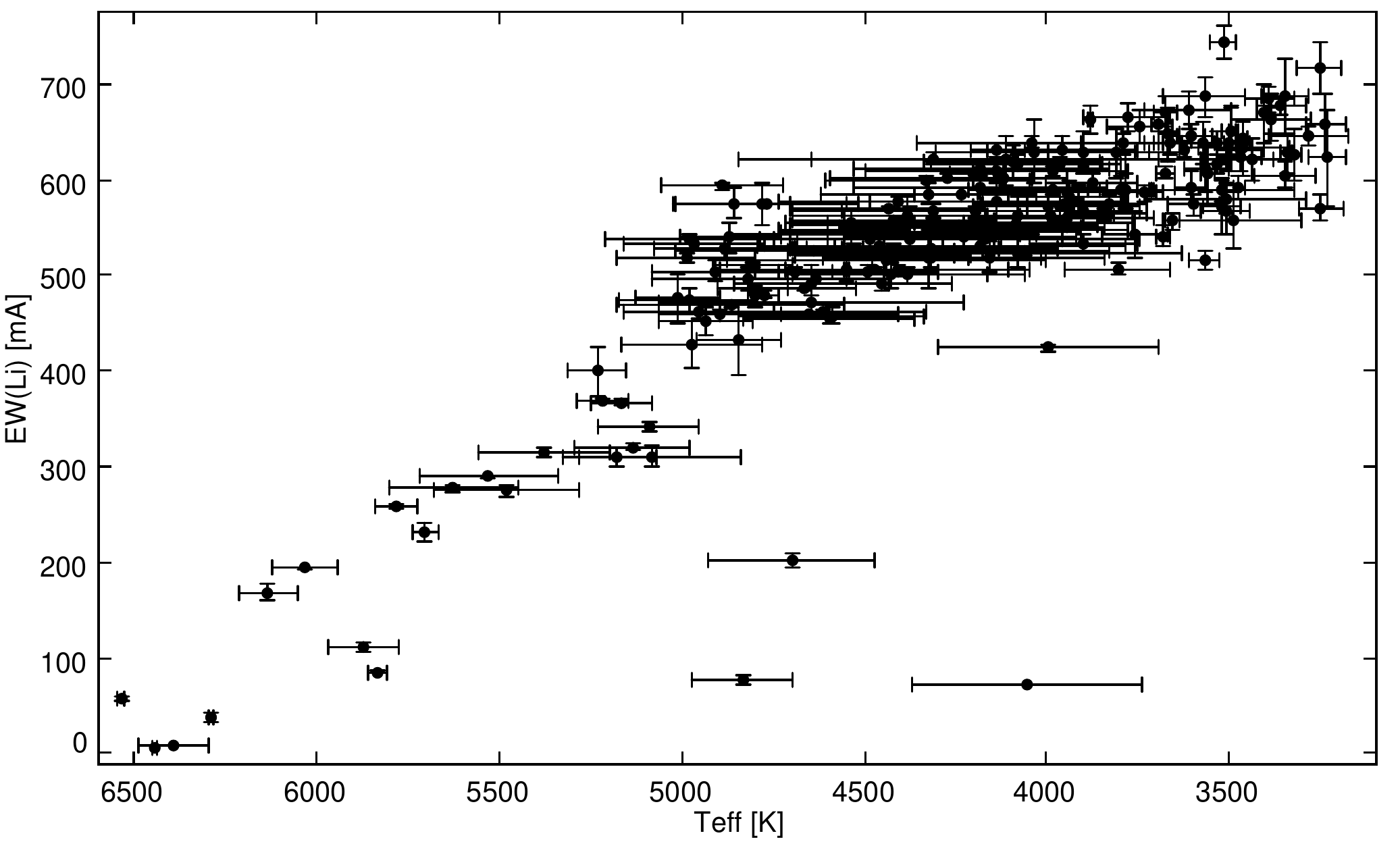}
   \caption{The lithium equivalent width
     is plotted as a function of effective temperature for weak-line T
     Tauri stars. The
     average rms error is 200~K on \teff\ and 10 m\AA\ on \ewli.} 
              \label{LiTeffWTTS}
    \end{figure}

Figure~\ref{LiTeffWTTS} shows \ewli\ plotted as a
function of effective temperature for the WTTS only. A strong dependency
is observed between \ewli\ and \teff, with \ewli\ rapidly increasing towards
lower \teff, as previously reported for various young clusters
\citep[e.g.][]{Bayo11, Jeffries14b}. This trend is also seen for CTTS (not shown) whose \ewli\ is,
however, affected by veiling, and therefore consistently lower at a
given effective temperature than those of WTTS. The limited accuracy of veiling measurements makes more uncertain the derivation of intrinsic, i.e., veiling-corrected,  \ewli\ for CTTS. We therefore elected to consider only WTTS for this study and thus concentrate on the 201 non-accreting
sources in the sample. As is apparent from Fig.~\ref{LiTeffWTTS}, nine of these have a \ewli\ 
significantly smaller than expected for their \teff. Of these, eight also
have discrepant radial velocities relative to the cluster's
average\footnote{These are (in order of decreasing \teff\ and with \vrad\ given in \kms\ within parentheses): Mon 1292 (28.3), 661 (0.0), 38 (13.6), 1282 (10.6), 560 (4.1), 351 (27.0), 411(27.1), 160 (31.7).} \citep[<\vrad>=19.8$\pm$2.5~\kms,][]{Jackson16}. We rejected those 9 sources as likely non-members\footnote{The  ninth source we reject is Mon 1115 (\teff =4833~K) which, in spite of its low \ewli=75 m\AA, has \vrad=18.3~\kms\ that is consistent with the cluster's average, as well as a high \vsini=30 \kms\ for its K6 spectral type, a sign of youth. It therefore deserves further membership analysis.} and  this leaves us with 192 WTTS to investigate further their lithium content. 




   \begin{figure}
   \centering
    \includegraphics[width=9cm]{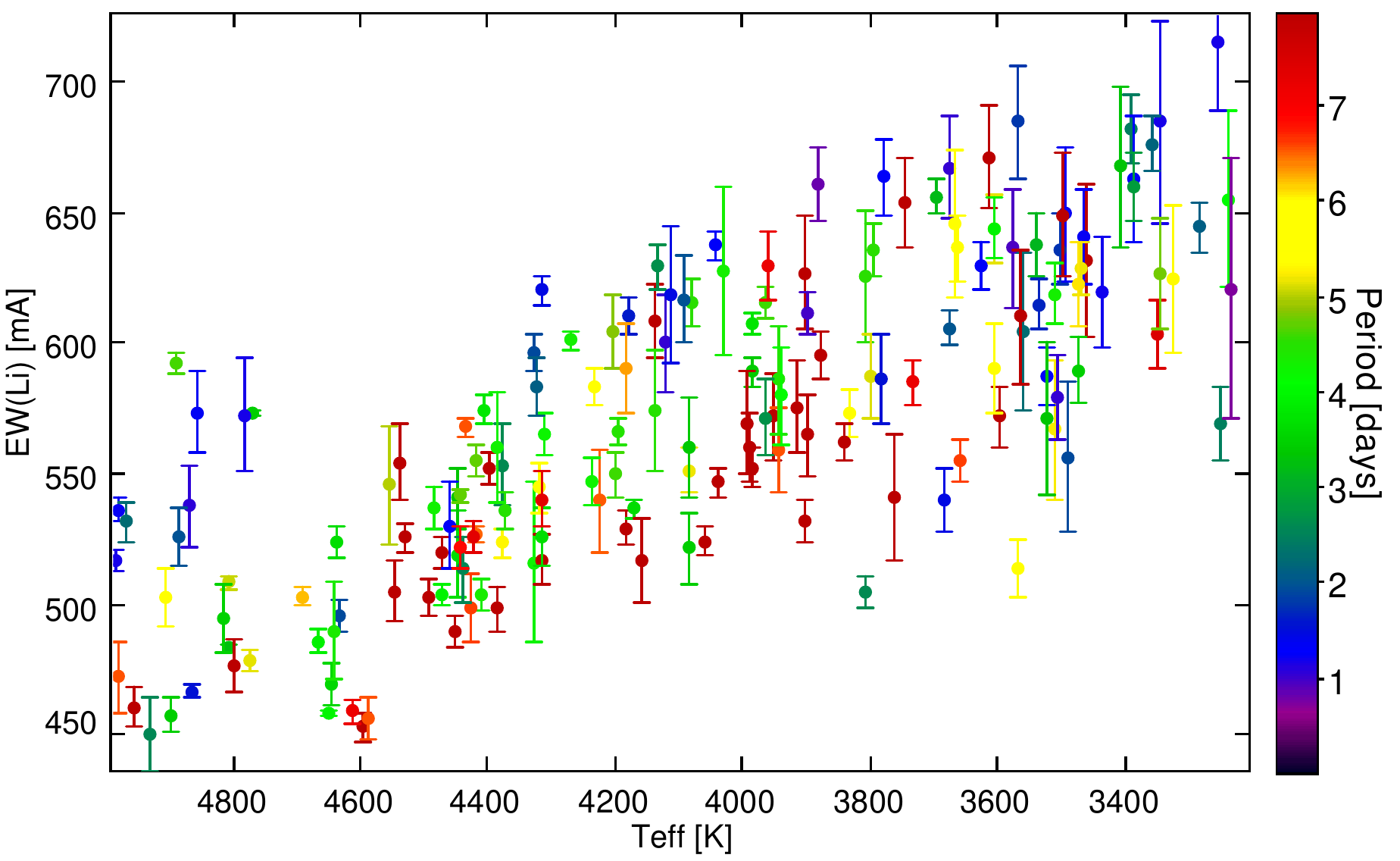}
    \caption{The lithium equivalent width
      is plotted as a function of effective temperature for weak-line
      T Tauri stars.  The decreasing \ewli\ trend towards higher \teff\ 
      continues up to 6600K (cf. Fig.~\ref{LiTeffWTTS}). The
      color scale is representative of the rotational period, and is limited up to 8~days in order to maximize the contrast (see the
      scale on the right side of the panel in units of days).}
              \label{LiTeff}
    \end{figure}

Figure~\ref{LiTeff} highlights the lower \teff\ range already shown in
Fig.~\ref{LiTeffWTTS}, from 3300~K to 5000~K.  The color code scales
with rotational period, which for stars in our sample ranges from less
than 1~day up to more than 12 days.  The
figure reveals significant lithium scatter over some
effective temperature ranges. This is most notably the case for
\teff$\leq$3800~K, i.e., M-type stars. \ewli\ is notoriously
difficult to measure in such cool stars, as the line is located in
deep TiO absorption bands. We therefore cannot exclude that the lithium
measurement error is underestimated for the very low mass stars in our
sample, and could at least in part account for the enhanced
Li scatter over this temperature range. 

Such an explanation, however, cannot hold for K stars where the
stellar continuum is relatively flat around the Li 6708\,\AA\ line. Its  
 equivalent width can thus be easily and accurately measured at
the spectral resolution of VLT/Flames (cf. Appendix~A).  Interestingly, over the temperature range from about
3800~K up to about 4400~K (SpT$\sim$K4--M0), Fig. ~\ref{LiTeff} shows
that fast rotators tend to lie on the upper part of the \ewli - \teff\ 
sequence, while many of the slow rotators lie on the lower part of
it. This suggests that the lithium scatter seen over this temperature
range could be linked with the rotational properties of the stars.  We refer the reader to Appendices A and B where we argue that this scatter does not result from a rotation-induced bias in the measurement of \ewli\ nor from intrinsic \ewli\ variability. We further note that this correlation is not seen at lower \teff, below 3800~K.

     \begin{figure}
   \centering
        \includegraphics[width=8cm, angle=0]{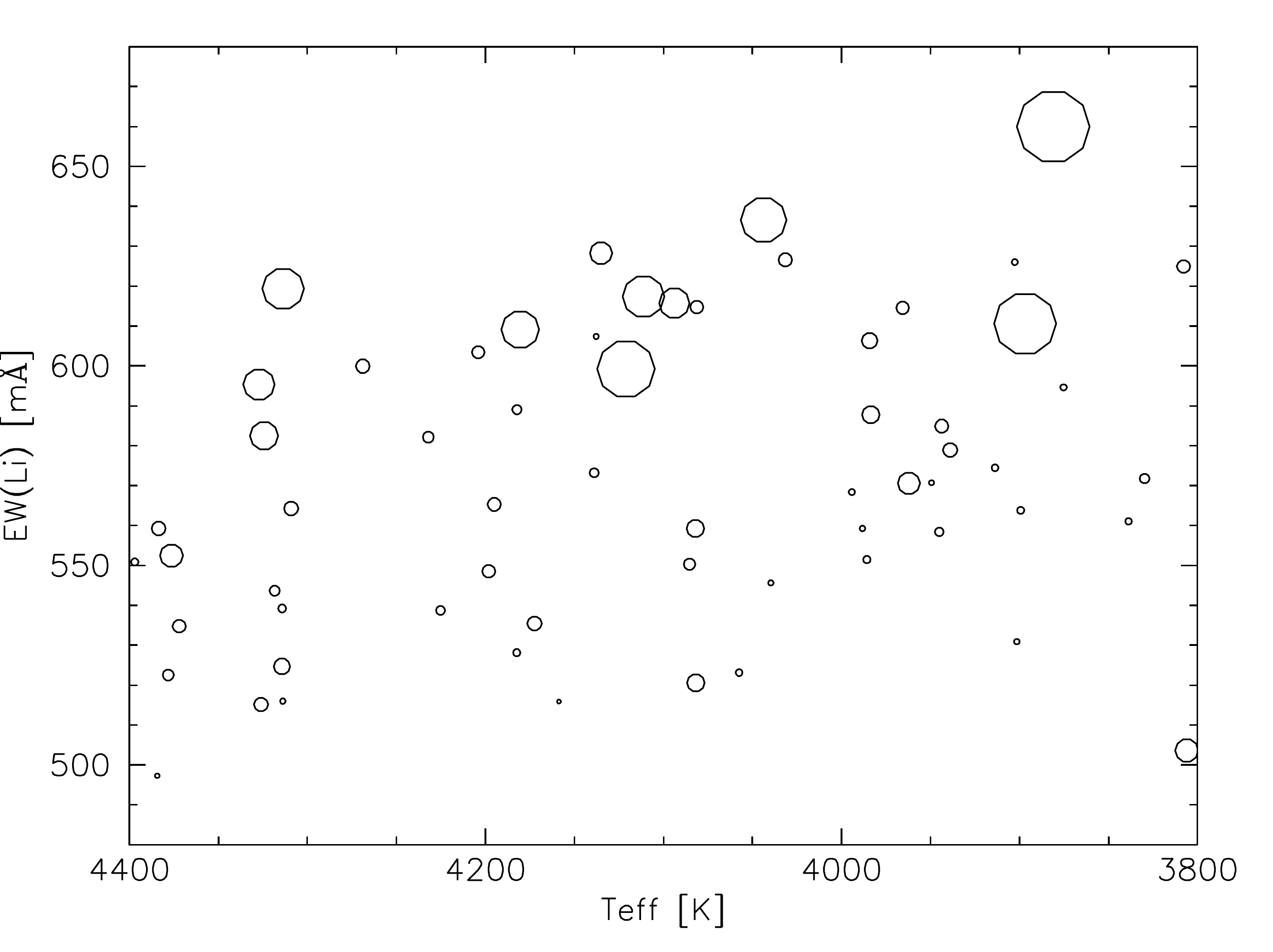}
   \caption{The lithium equivalent width
     is plotted as a function of effective temperature for weak-line T
     Tauri stars over the \teff\ range 3800--4400K. The symbol size is
     proportional to the angular velocity ($1/{\rm P}_{rot}$). } 
              \label{LiTeffProt}
    \end{figure}
  
    In order to put the above statement on a firmer quantitative
    ground, we proceeded as follows. Figure~\ref{LiTeffProt} shows \ewli\ 
    as a function of \teff\ over the 3800--4400~K range with a symbol size
    proportional to the angular velocity, $\Omega\propto 1/{\rm P}_{rot}$. This subsample includes 62 WTTS with accurate \ewli\ measurements and robust period determinations\footnote{We discarded two stars in this \teff\ range, namely Mon~1015 and Mon~1236, as they exhibit multiple periods.}. A
    similar figure is obtained when the symbol size is taken to be
    proportional to ${\rm V}_{eq} = 2\pi {\rm R}_\star/{\rm P_{rot}}$, where R$_\star$
    is the stellar radius taken from \citet{Venuti14}. The trend of increasing \ewli\ towards lower
    \teff\ is clearly seen (cf. also Fig.~\ref{LiTeffWTTS}) and we assume
    it is linear in the restricted \teff\ range considered here to 
    obtain: 
       $$ \rm{EW}(\rm{Li})_{lsq} / \mathrm{m\AA} = - \mathrm{T}_{eff} / 14.41\mathrm{K} + 857 \mathrm{m\AA} $$
        from a linear least-square
    fit. We then compute the \ewli\ residuals relative to the fit as:
    $$ \delta \rm{EW}(\rm{Li}) = \rm{EW}(\rm{Li}) - \rm{EW}(\rm{Li})_{lsq}$$ In the following, we refer
    to ``Li
    excess'' stars as those that have a positive \dewli, and to ``Li
    deficient'' stars as those with a negative \dewli.  


     \begin{figure}
   \centering
      \includegraphics[width=8cm]{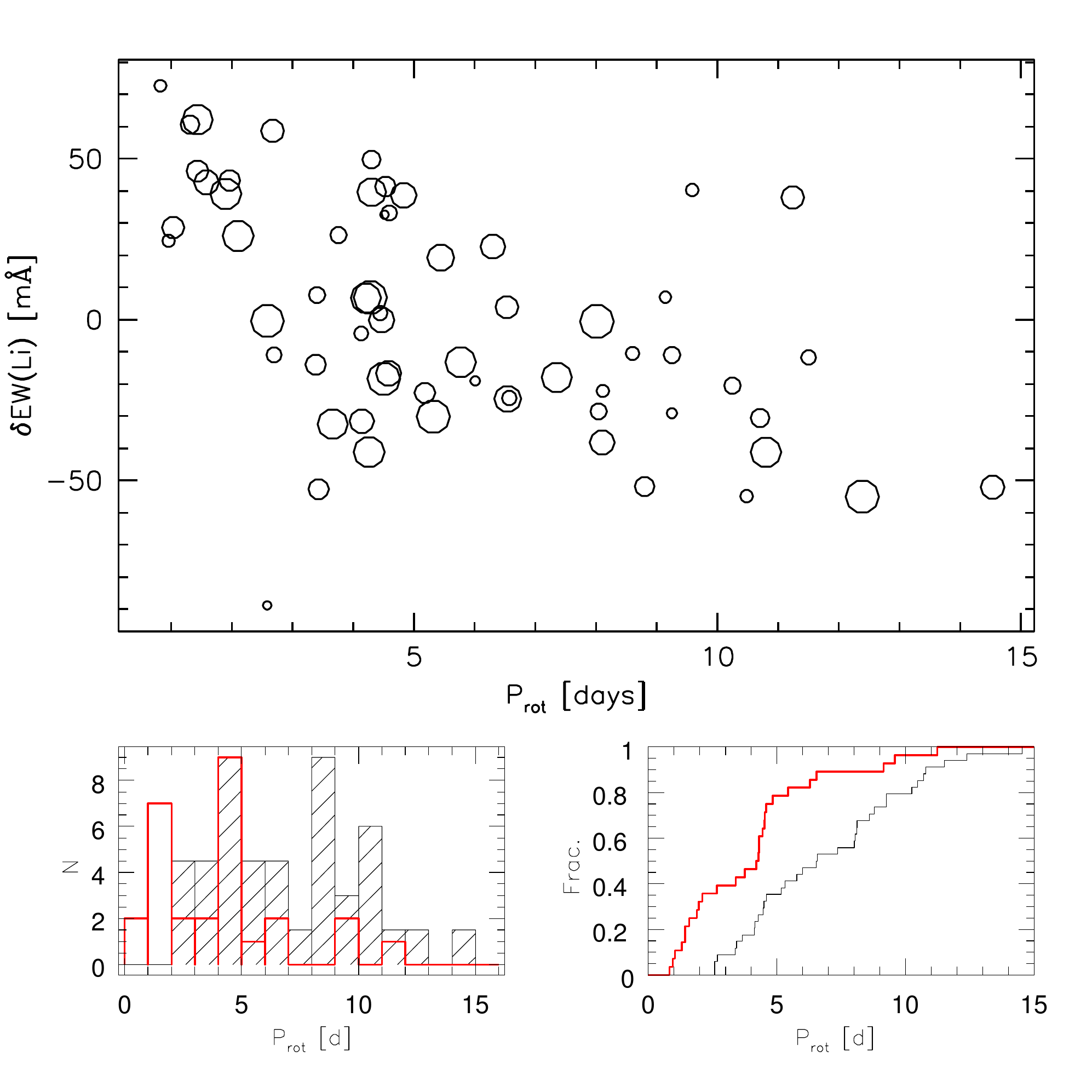}
   \caption{{\it Top panel:} The \ewli\ 
     residuals are plotted as a function of rotational period. The
     symbol size is proportional to \teff. {\it Lower left panel:}
     Rotational period distribution of Li-excess (thick red) and Li-deficient
     (shaded black) stars. {\it Lower right panel:} Cumulative
     distribution of rotational periods for Li-excess (thick red) and
     Li-deficient (thin black) stars.  } 
              \label{LiProtKS}
    \end{figure}

    Figure~\ref{LiProtKS} shows \dewli\ plotted as a function of
    rotational period. An overall trend is seen for the Li excess to
    be preferentially associated to fast rotators (short rotational
    periods) while Li-deficient stars appear to be present at all but
    the shortest periods. Histograms of the rotational distributions
    of Li-excess and Li-deficient stars are also shown in
    Fig~\ref{LiProtKS}. The Li-deficient stars exhibit a broad distribution over the 2--11~d period range, while the Li-excess stars appear to concentrate at periods shorter than 5~d.  
     As a check, a two-sided
    Kolmogorov-Smirnov test was run on the rotational distributions of
    Li-excess and Li-deficient stars. The test returns a probabilty of
    $4\times10^{-3}$ for the two samples to be drawn from the same parental
    population\footnote{A Spearman rank test  \citep{Press92} yields a correlation coefficient of $\rho=-0.50$  with a
significance of level $\sigma=2.9\times10^{-5}$, that indicates a highly significant
correlation between \dewli\ and P$_{rot}$.}\fnmsep\footnote{The outlier in Figs.~\ref{LiTeffProt} and \ref{LiProtKS} is Mon 910 (\teff=3806~K, \ewli=503.6 m\AA, ${\rm P}_{rot}$=2.58~d). Its \ewli\ is lower than that of other cluster members of the same \teff, which might indicate it is a slightly older star unrelated to the cluster. Besides, it lies at the edge of the temperature domain considered here. We checked that excluding this object from the sample does not affect the significance results (the KS probability becomes $7\times 10^{-3}$), although it does change the \ewli--\teff\ least-square fit and thus the actual \dewli\ values.}.




\section{Discussion}



The results reported above suggest that a relationship exists between the lithium content and the spin rate of young stars at an age as early as 5~Myr, with faster rotators being more lithium rich. The lithium-rotation connection is best seen for NGC 2264 members spanning the \teff\ range from about 3800~K to 4400~K, which corresponds to a mass range of 0.5--1.2~\msun\ at that age \citep{Siess00}. This encompasses the mass range over which a similar lithium-rotation relationship has previously been reported for the 125~Myr-old Pleiades dwarfs, with more rapidly rotating K stars being more lithium rich than slow rotating ones (Soderblom et al. 1993). A variety of ideas have been put forward to account for the lithium-rotation connection at the age of the Pleiades on the ZAMS. In this section, we examine whether any of the proposed scenarios is consistent with the appearance of the Li-rotation relationship indeed occuring much earlier in time during the PMS, or whether alternative explanations have to be sought for. 

The lithium equivalent width of fast rotators in NGC~2264 is of order of 600 m\AA\ over the \teff\ range from 4000 to 4400~K (cf. Fig.~\ref{LiTeffProt}). According to the Local Thermodynamic Equilibrium (LTE) curves-of-growth published by  \cite{Soderblom93},  and in agreement with lithium abundances reported for these stars in the GES internal data release, this corresponds to a lithium abundance of A[Li]$\simeq$3.0--3.2 dex\footnote{Non LTE corrections depend on the temperature, gravity, and Li abundance itself \citep[see e.g.,][and refs. therein]{Lind09}, but, generally, for cool stars at solar metallicity the corrections are small ($\leq$0.1~dex) and negative, i.e., Non LTE abundances are slightly lower than LTE ones. Hence the maximum abundances measured in NGC 2264 would be consistent with the meteoritic abundance.}. This suggests that fast  rotators have almost no lithium depletion at the age of NGC~2264 compared to the  meteoritic abundance \citep{Soderblom99}. The lower \ewli\ of slow rotators over the same temperature range, amounting to around 530~m\AA\ on average, translates to a lithium abundance of A[Li]=2.8--3.0 dex, i.e., about 0.2 dex less than that of fast rotators. The models by \cite{Baraffe15}  indeed predict a 0.12--0.25~dex lithium depletion relative to initial abundances for stars over this \teff\ range at 5 Myr. Over the same \teff\ range, the models by \cite{Siess00} also predict significant lithium burning, with $\delta$[Li]$\sim$0.05--0.20 dex at this age. Assuming an age of 5~Myr for the cluster \citep{Gillen14}, these models would then suggest that slow rotators behave as expected, while fast rotators have apparently been prevented from depleting lithium. However, assuming an age of 3 Myr for the cluster \citep{Mayne08}, the same models would suggest instead that fast rotators are Li-undepleted as expected for this younger age, while slow rotators have undergone enhanced lithium depletion. Moreover, PMS lithium depletion, as predicted by standard  evolutionary models, is highly sensitive to the physics included in the model, most notably the treatment of convection \citep[e.g.,][]{Jeffries14}. Hence, any model-based prediction of lithium abundances as a function of mass and age in the PMS is subject to those uncertainties, as well as being quite sensitive to the assumed cluster age. 

An immediate interpretation of the lithium scatter would be to assign it to a corresponding age spread among the stars in our subsample \citep[e.g.][]{Palla05, Zwintz14}. According to the models by \cite{Baraffe15} the required age spread would be of the order of 2--3 Myr, a significant fraction of the cluster age. It is notoriously difficult to measure individual ages for young stars \citep[e.g.][]{Soderblom14}. Nevertheless, following \cite{Venuti14}, we  checked \ewli\ against the age values they report for our subsample, as well as the location of the same stars in the HR diagram and their spatial distribution over the star forming region. None of these checks provided any evidence for an age spread being responsible for the observed Li scatter. In addition, it could be expected that the older WTTS have spun up to faster rates than the younger ones. This would then result in a lower lithium content in faster rotators if only an age effect, a trend that is opposite to the lithium-rotation relationship reported here. Hence, a significant age spread among the cluster members would tend to minimize the observed lithium spread. Although larger samples and better individual age estimates are clearly needed to fully tackle this issue, we turn to discuss alternative scenarios below. 

\cite{Soderblom93} reported lithium measurements for about 100 F, G, and K dwarfs in the Pleiades. They confirmed the large lithium abundance spread, most notably seen in K dwarfs where it amounts to 1--1.5 dex. Within this spread, they showed that the fastest rotators showed the largest lithium abundances \citep[see also][]{Butler87}. They investigated effects that could possibly impact on the lithium abundance determination, such as chromospheric activity and/or stellar spots, and concluded that they cannot account for their results \citep[see also][]{Favata95, King10}. They thus claimed that the connection between excess Li and rapid rotation is real and clearly present among Pleiades low-mass stars on the ZAMS. While this early result raised fundamental issues regarding the PMS lithium depletion history of cool dwarfs and its relationship with angular momentum evolution, no satisfactory explanation could be offered at that time. 

Since then, \cite{Baraffe10} revisited the issue of PMS lithium depletion in the framework of episodic accretion. They showed that repeated accretion bursts during the early PMS can modify the internal structure of young low-mass stars in such a way as to enhance their central temperature, hence increasing the lithium depletion rate. Indeed, their model predicts that episodic accretion onto an initial protostellar seed of 10 Jupiter masses, which will eventually produce a solar-mass star, would fully deplete lithium in less than 1 Myr. In contrast, standard non accreting models would preserve most of the initial lithium content for several Myr \citep[e.g.][]{Siess00}. Hence,  a coeval young stellar population may exhibit a significant dispersion in lithium content at a few Myr depending of the specific accretion history of its members. As noted in \cite{Baraffe10}, this model prediction heavily relies on the assumptions of a very low-mass protostellar seed combined with a low specific entropy for the accreted material. For a larger initial seed and/or a higher fraction of internal accretion energy being absorbed by the protostar, the lithium depletion rate is similar to that of non-accreting models. Indeed, these assumptions are challenged by the lack of a significant scatter observed in the lithium abundances of the accreting populations of the Orion Nebula Cluster and NGC 2264 \citep{Sergison13}. We cannot totally exclude a different accretion history for the subsample of NGC 2264 low-mass members investigated here. However, the lithium-rotation link reported here would additionally call for a direct impact of episodic accretion onto angular momentum evolution. More specifically, in order to recover the enhanced lithium depletion rate in  slowly rotating stars, one would have to assume that accretion bursts also yield lower spin rates at a few Myr. This is somewhat paradoxical, as violent accretion bursts are expected to spin the central star up, though it must be acknowledged that the spin evolution of young eruptive stars remains uncertain \citep[e.g.][]{Popham96}.    

A more direct link between PMS lithium depletion and angular momentum evolution has been suggested by \cite{Bouvier08}. Recent models of PMS spin evolution indicate that the outer convective envelope of contracting young stars is held at nearly constant angular velocity for a few Myr through its magnetic coupling with the circumstellar disk \citep[e.g.][]{Gallet13}. As the inner radiative core develops, unimpeded by the external braking torques, differential rotational sets in between the fast radiative core and the slower outer convective envelope. In turn, the internal angular velocity gradient promotes increased rotational mixing, which results in an enhanced lithium depletion. The models predict that slow rotators remain coupled to their disk for a longer period of time and therefore experience a larger differential rotation than fast rotators \citep[cf.][]{Gallet15}. The former are thus expected to reach the ZAMS with a lower lithium content than the latter, as observed in the Pleiades. Whether this process is able to account for the development of the lithium-rotation connection at an age as early as 5~Myr is, however, unclear. Non accreting stellar evolution models predict that a 0.8~\msun\ star starts to develop a radiative core at about 3 Myr \citep{Siess00, Baraffe15}. At 5 Myr, the models by \cite{Baraffe15} predict that stars less massive than 0.6~\msun\ are still fully convective while stars in the mass range from 0.7 to 1.0~\msun\ have a radiative core whose mass increases from 5 to 30\%, respectively, of the stellar mass, and whose radius extends from about 20 to 40\% of the stellar radius\footnote{Similar figures arise from models by \cite{Siess00}.}. Hence, over the \teff\ range investigated here, only the more massive WTTS may possess a significant radiative core at NGC 2264's assumed age. It therefore remains to be seen whether a prolonged disk braking phase lasting a few Myr may effectively produce the level of lithium dispersion observed here for NGC~2264 low-mass members on a timescale as short as 5~Myr \citep[see][]{Eggenberger12}\footnote{P. Eggenberger kindly ran a 0.8~\msun\ rotating model with the same assumptions as the 1~\msun\ model published in \cite{Eggenberger12} and obtained similar results, namely an insignificant impact of enhanced rotational mixing upon lithium depletion at an age of 5 Myr.}. Nevertheless, we note that the amplitude of the lithium scatter increases from a mere 20\% in NGC~2264 (see Sect.~2) to a factor of 6 in the Pleiades \citep{Soderblom93}. Hence, the long-lasting core-envelope decoupling process triggered by extended disk lifetime in the early PMS may still play a central role during the late PMS stage in widening the initially modest lithium dispersion up to the ZAMS.




A more subtle link between rotational evolution and lithium depletion during the PMS has been recently advocated by \cite{Somers14}, based on an earlier suggestion by \cite{King10}. Following models by, e.g.,  \cite{Chabrier07}, they argue that the strong magnetic fields of young stars partly inhibit the convective energy transport and therefore lead to inflated radii. In turn, an enhanced radius results in a lower temperature at the bottom of the convective zone and therefore reduces the rate of lithium burning. They further argue that, {\it if} magnetic field strength scales with rotation in young low-mass stars, fast rotators will have the most inflated radius and therefore will experience only reduced lithium burning. Hence, the Li excess observed in fast rotators by the Pleiades age would ultimately result from the impact of dynamo-generated magnetic fields scaling with rotation rate onto the PMS stellar structure. The models by \cite{Somers14} predict a slight depletion of lithium abundances already occurring at 6~Myr over the \teff\ range investigated here, with a dispersion of about 0.4 dex, i.e., of similar magnitude as or even slightly larger than observed here for NGC 2264 (see their Fig.~18). We note, however, that the assumption of a magnetic field strength scaling with rotation rate might not hold during the PMS. While this relationship is well established for main sequence dwarfs \citep[e.g.][]{Petit08, Donati09}, PMS stars, slow and fast rotators alike, appear to lie in a ``saturated'' magnetic activity regime at low Rossby numbers \citep{Gregory12, Donati13}. Hence, even though PMS low-mass stars may have an inflated radius due to enhanced magnetic activity compared to mature dwarfs \citep{Jackson16}, the relationship between radius inflation, magnetic strength, and spin rate during the PMS, a central assumption indeed for this scenario to hold is still a matter of debate \citep[see][]{Somers15}. We empirically tested whether we could find in our subsample of WTTS any correlation between X-ray luminosity and rotation (E. Flaccomio, priv. comm.), or between amplitude of variability and rotation (Venuti et al., in prep.) and found none that would support a clear relationship between rotation and either coronal emission or 
spottedness in this subsample \citep[see also][]{Messina03, Argiroffi16}. 
 


Finally, since none of the above scenarios is totally exempt from difficulties in explaining how a lithium-rotation connection may appear in low-mass PMS stars at an age of 5~Myr, we have to contemplate other possibilities and envision in particular factors extrinsic to the stars, such as accretion of fresh lithium from the circumstellar disk or through planet engulfment. We note that the age of the NGC~2264 cluster is similar to the average lifetime of accretion disks around young low-mass stars \citep{Bell13}. Hence, a fraction of stars in our sample must have lost their disk quite recently. According to \cite{Vasconcelos15} Monte Carlo simulations, non accreting stars at 5 Myr span the whole range of rotational periods from less than 2 days up to 15 days. Slowly rotating non accreting stars have lost their disk only recently and had no time to spin up yet, while fast rotating WTTS have lost their disk earlier and have already had time to spin up under contraction. Fast rotating WTTS are thus more evolved than slow rotating ones, at least in terms of disk evolution. If fresh lithium can be provided to the star from the disk during the steady accretion phase while the star is still fully convective, one would then expect slowly rotating WTTS that had a prolonged disk accretion phase to exhibit higher lithium content\footnote{In contrast, when accretion of heavy material occurs once the star has settled on the ZAMS, it promotes lithium depletion \citep[cf.][]{Theado12}.}. This is opposite to what is actually observed.  Alternatively, if the disk dissipation is somehow linked to planet formation, young planetary systems around WTTS might already be in the process of early dynamical evolution \citep[e.g.][]{Albrecht12}. Random gravitational encounters may then promote dynamical ejections and, eventually, planet engulfment by the central star that would then gain angular momentum \citep{Bolmont12, Bouvier15} and be replenished in lithium, albeit at a very modest rate for fully convective stars unless a significant amount of Li-rich planetary material is accreted. Admittedly, this interpretation is bound to remain speculative until we better understand how planetary systems form and evolve around young stars. 






\section{Conclusions}

We report here a connection between lithium content and rotation rate for low-mass members of the 5~Myr-old NGC 2264 cluster over a restricted \teff\ range from 3800~K to 4400~K. The amplitude of the \ewli\  scatter over this \teff\ range is about 20\%, i.e., much smaller than the lithium dispersion reported for K-dwarfs in the 125~Myr-old Pleiades cluster, which amounts to a factor of six. The relationship between lithium content and rotation nevertheless goes in the same direction, with fast rotators being more Li-rich than slow ones. This strongly suggests that a lithium-rotation connection is already established early on during PMS evolution, and continues to grow further up to the ZAMS. Among the non-standard evolutionary models developed so far to account for a lithium-rotation connection in low-mass dwarfs, those calling for a reduced lithium depletion rate during the PMS, as a result of radius inflation linked to fast rotation and magnetic activity, may prove the most able to account for the results reported here. Yet, the long-lasting impact of disk lifetime on the internal rotation profile of young solar-type stars may also contribute to the large lithium scatter observed on the ZAMS and beyond. 

None of these models is without difficulties and external factors cannot be excluded at this point, such as the prompt injection of fresh lithium to the star by either disk accretion or planet engulfment. The evidence we report here for a rapidly increasing lithium scatter as a function of age, from the early PMS to the ZAMS, is quite reminiscent of the observed widening of the angular momentum distribution of low-mass stars during this phase. Hence, the existence of a lithium-rotation connection in low-mass stars at 5 Myr may be indicative of an intimate link between accretion history, planet formation, and early angular momentum evolution, a complex interplay indeed whose breadth and underlying physics remain to be fully deciphered. As additional lithium measurements will become available for other young clusters from the \textsl{Gaia}-ESO Survey, and rotational period distributions will be derived from continuous photometric monitoring, it will be most interesting to trace the origin and development of the lithium scatter and its relationship with rotation along PMS evolution, from a few Myr up to the ZAMS.

%

\begin{acknowledgements}
We would like to dedicate this paper to our friend and colleague Francesco Palla, a pioneer in the investigation of lithium in young stars. We thank Patrick Eggenberger for running additional models of PMS lithium depletion, and Silvano Desidera and Elvira Covino for sharing results on lithium
rotational variability in advance of publication. J. Bouvier and E. Moraux thank the staff of the Osservatorio Astrofisico di Catania for their kind hospitality in automn of 2015. They acknowledge the Agence Nationale pour la Recherche program ANR 2010 JCJC 0501 1``DESC (Dynamical Evolution of Stellar Clusters)'' for the funding of their stay at OACT during which this study was performed. This study was also supported by  the Agence Nationale pour la Recherche program ANR 2011 Blanc SIMI 5-6 020 01 ``Toupies (Towards understanding the spin evolution of stars )''. A. Bayo acknowledges financial support from the Proyecto Fondecyt Iniciaci\'on 11140572. S.G. Sousa acknowledges the support from FCT through Investigador FCT contract of reference IF/00028/2014. E. Delgado Mena acknowledges the support from FCT in the form of the grant SFRH/BPD/76606/2011. S.G.S. and E.D.M. also acknowledge the support from FCT through the project PTDC/FIS-AST/7073/2014

Based on data products from observations made with ESO Telescopes at the La Silla Paranal Observatory under programme ID 188.B-3002. These data products have been processed by the Cambridge Astronomy Survey Unit (CASU) at the Institute of Astronomy, University of Cambridge, and by the FLAMES/UVES reduction team at INAF/Osservatorio Astrofisico di Arcetri. These data have been obtained from the \textsl{Gaia}-ESO Survey Data Archive, prepared and hosted by the Wide Field Astronomy Unit, Institute for Astronomy, University of Edinburgh, which is funded by the UK Science and Technology Facilities Council. This work was partly supported by the European Union FP7 programme through ERC grant number 320360 and by the Leverhulme Trust through grant RPG-2012-541. We acknowledge the support from INAF and Ministero dell' Istruzione, dell' Universit\`a' e della Ricerca (MIUR) in the form of the grant ``Premiale VLT 2012'' and ``The Chemical and Dynamical Evolution of the Milky Way and the Local Group Galaxies" (prot. 2010LY5N2T). The results presented here benefit from discussions held during the \textsl{Gaia}-ESO workshops and conferences supported by the ESF (European Science Foundation) through the GREAT Research Network Programme.

\end{acknowledgements}

\bibliographystyle{aa} 
\bibliography{li_rev3} 

\begin{appendix}
\section{Lithium equivalent width measurements in NGC~2264}

As described in \cite{Lanzafame15}, the \textsl{Gaia}-ESO PMS analysis makes use of three independent methods to derive \wli\ from the GIRAFFE spectra\footnote{These methods are: the direct profile integration available in the \texttt{IRAF-splot} procedure (OACT node); DAOSPEC \citep[][OAPA node]{2008PASP..120.1332S}; and a semi-automatic IDL procedure specifically developed for the \textsl{Gaia}-ESO by the Arcetri node.}.
In order to improve further the reliability of the final recommended \wli, a robust average procedure was introduced in iDR4 for deriving the final recommended values from the independent measurements\footnote{In iDR4 at least two of the independent measurements must be available before combining the results into the final recommended value (A.C. Lanzafame, priv. comm.). When three measurements are available, the outlier-resistant mean procedure \texttt{resistant\_mean} in the IDL Astronomy Library (\url{http://idlastro.gsfc.nasa.gov}) with 3$\sigma$ outlier rejection is used. 
When two measurements are available, they are averaged only if they differ by less then 2$\bar{\sigma}$, with $\bar{\sigma}$ being the mean of the uncertainties of the individual measurements, otherwise no final recommended value is given.}
Hence, the final recommended \wli\ for NGC~2264 have a median standard deviation of 10\,m\AA, with 99\% of the values having an estimated precision better than 6\%. We further checked this estimate by analyzing a subsample of 12 stars over the \teff\ range 3800-4440~K that have been observed repeatedly on a timescale of several weeks. Each of these sources has at least 20 FLAMES spectra in the ESO archive and \ewli\ was measured on each spectrum. The results indicate an average rms dispersion around the mean of 5--6 m\AA, in agreement with the above estimate.  

The contribution of lines blended with \li\ in the GIRAFFE spectra is estimated by a spectral synthesis using Spectroscopy Made Easy \citep[SME,][]{1996A&AS..118..595V} with MARCS model atmospheres as input, taking  the stellar \teff, \logg, and \feh\ into account. Blends are estimated on a wavelength range of 1\,\AA\ centered on the \li\ line, then subtracted from the raw \wli.
Since \vsini\ is not taken into account as yet, further lines can in principle contribute to the measured \wli\ for sufficiently fast rotating stars, which can introduce a bias.
In order to estimate such a bias at increasing \vsini, we have taken a dozen HR15N spectra of young stars with S/N$\approx$60, 400$\le$\wli$\le$500m\AA, 3800$\le$\teff$\le$4400\,K, and \vsini$<$20~\kms, convolved them with rotational broadening kernels to simulate the effect of increasing rotation velocity, measured \wli\ in each of them, and compared the results with the original spectra.
The results are summarized in Fig.\,\ref{fig:EWLI_broaden_spectra}.
Differences with the original spectrum, $\Delta$\wli, remain less than the median standard deviation in NGC~2264 up to \vsini$\approx$80~\kms\ and become increasingly significant only above this value, where we have only five NGC~2264 members in the sample considered here.
Furthermore, we estimate that only the presence of a significant fraction of stars with \vsini$>$100\,\kms\ would introduce a bias in the observed distribution of \wli\ vs. rotation, but indeed only one NGC~2264 member in our sample lies in this range.
Fig.\,\ref{fig:EWLI_broaden_spectra} also shows that, in some cases, for 20$<$\vsini$<$50\,\kms\  the \wli\ measurement can actually be slightly less than the original measurement, but still within the typical standard deviation, since the merging of lines due to the rotational broadening can lead to a small underestimation of the local continuum.   

We therefore conclude that the systematic differences in the rotation distribution of Li-excess and Li-deficient stars in NGC~2264 cannot be due to such a bias.

\begin{figure}[htp]
\centering
\includegraphics[width=8cm]{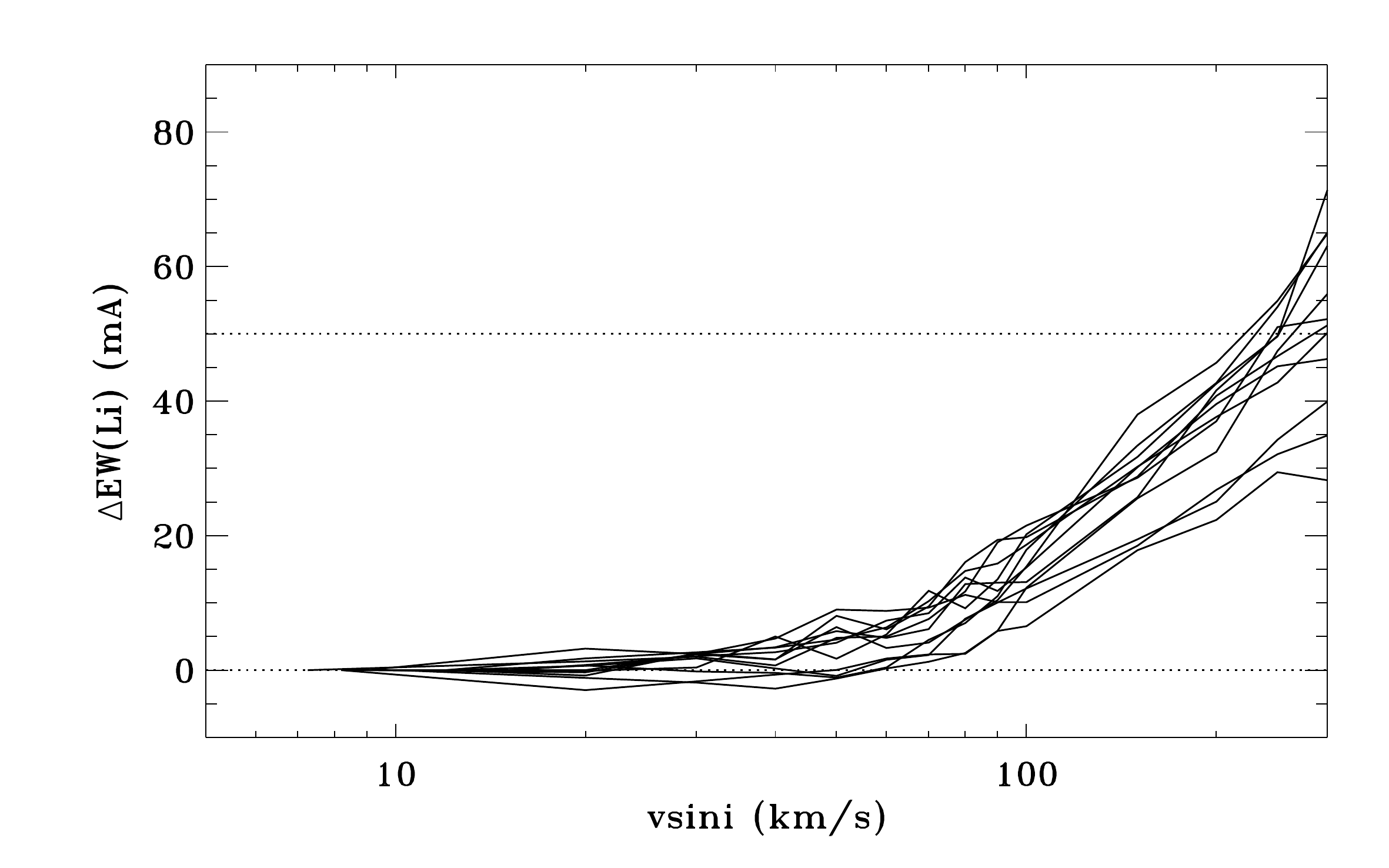}
\caption{Differences in \wli\ measurements between artificially broadened spectra and the original spectra for a selected sample of stars (see text for details). The dotted line marks the approximate distance between the Li-excess and Li-deficient distributions.}
\label{fig:EWLI_broaden_spectra}
\end{figure}

\section{Lithium equivalent width intrinsic variability} 

We investigate here how intrinsic \ewli\ variability could impact our results. The first evidence for the rotational modulation \ewli\ 
on spotted low-mass stars was reported by \cite{Robinson86}. They  
observed an increased \ewli\ when
the star was fainter, i.e., at maximum spot visibility. The \ewli\  enhancement in spotted areas  
was already well documented in the 
case of the Sun \citep{Giampapa84}. However, there are also cases when the \ewli\ dependence
on the spottedness level is observed only at some epochs,
e.g. V410 Tau \citep{Fernandez98}, but is absent at other epochs 
\citep[see][]{Basri91}. Also, spectroscopic monitoring of some active stars in the Pleiades revealed no  significant Li variations \citep{Pallavicini93, Jeffries99}. 
This non univocal behavior suggests that additional factors other than spots
can play a role in the observed \ewli\ variations.

The amplitude of the \ewli\  rotational modulation is generally of a few percent.
Messina et al. (in prep.) measured \ewli\ variations in a sample 
of 18 low-mass stars over the age range from 6~Myr to the Pleiades age. The \ewli\ variations
were measured in all stars, over up to three conscutive years.
When present, \ewli\ variations were found to correlate with both
photometric variations and chromospheric activity indices (e.g., Ca II H\&K), with 
enhanced \ewli\ corresponding to light curve minimum 
and maximum chromospheric intensity. None of these \ewli\ variations exceeded 5\%. We therefore conclude that, while the NGC~2264 low-mass members may well undergo intrinsic \ewli\ variations, the increased lithium scatter to be expected remains small compared to the amplitude of the lithium-rotation connection reported here. Hence, intrinsic \ewli\ variations are unlikely to affect our results.

\end{appendix}

\begin{sidewaystable*}
\caption{NGC~2264 WTTS considered in this study. }             
\label{table1}      
\centering                          
\begin{tabular}{l l l l l l l l l l l l l l l l}        
\hline\hline                 
CSI-Mon   &   GES-NAME   &   RA(2000)   &   Dec(2000)   &   \teff   &   $\sigma$\teff & \ewli   &   $\sigma$\ewli   &   Flag$^\dagger$   &   Period   &   ref$^\ddagger$   &   \vrad   &   $\sigma$\vrad   &   \vsini   &   $\sigma$\vsini & Notes\\
 & & deg. & deg. & K & K &m\AA & m\AA & & days & & \kms & \kms & \kms & \kms & \\
\hline                        
000018   &   06411322+0955087   &   100.3052   &   9.91909   &   4326   & 114 &  515.1   &   30.4  &   1   &   4.26   &   V+16   &   17.7   &   0.9   &   30.0   &   0.5 \\
000020   &   06420924+0944034   &   100.53849   &   9.73427   &   4085   & 221  & 550.2   &   8.4   &   1   &   5.179   &   V+16   &   16.6   &   0.7   &   15.9   &   1.6 \\
000029   &   06410328+0957549   &   100.26367   &   9.96528   &   4396   &  307 & 550.8   &   6.1   &   1   &   8.012   &   V+16   &   19.1   &   0.7   &   15.0   &   0.5 \\
000033   &   06410726+0958311   &   100.28026   &   9.97533   &   4376   &  224 & 552.4   &   16.0   &   1   &   2.586   &   V+16   &   21.5   &   1.6   &   34.7   &   1.7 \\
000038   &   06411088+1000409   &   100.29532   &   10.01136   &   6289   &  7  & 34.8  &   4.3   &   1   &   3.615   &   V+16   &   13.6   &   0.7   &   24.0   &   0.5  & (1)\\
\hline                                   
\end{tabular}
\tablefoot{Only the first five lines of Table~1 are shown here. The full table is available electronically. \\
$^\dagger$ 1: \ewli\ is corrected from line blends contribution using stellar models (see text); 2: \ewli\ is measured separately from line blends (UVES spectra only); 3: Only an \ewli\ upper limit could be derived.\\
$^\ddagger$ V+16: Venuti et al. (in prep.); L+04: \cite{Lamm04}; M+04: \cite{Makidon04}; A+13: \cite{Affer13}\\
Notes: (1) This source has low \ewli\ for its \teff\ and a discrepant \vrad\ (see text); (2) This source has low \ewli\ for its \teff\ (see text); (3) This source has multiple periods detected in its light-curve (see text).}\\
\end{sidewaystable*}

\end{document}